\documentclass[twocolumn,epjc3]{svjour3}          % twocolumn

\RequirePackage[T1]{fontenc}

\smartqed  % flush right qed marks, e.g. at end of proof

\RequirePackage{graphicx}
\RequirePackage{mathptmx}      % use Times fonts if available on your TeX system
\RequirePackage{flushend}
\RequirePackage{amsfonts}
\RequirePackage[numbers,sort&compress]{natbib}
\RequirePackage[colorlinks,citecolor=blue,urlcolor=blue,linkcolor=blue]{hyperref}
\RequirePackage{amsmath}
\RequirePackage{epsfig}
\RequirePackage{bm}
\journalname{Eur. Phys. J. C}

\begin{document}

\title{Null paths on a toroidal topological black hole in conformal Weyl gravity}

\author{J. R. Villanueva\thanksref{e1,addr1}
	    \and
	    Francisco Tapia\thanksref{e2,addr1} 
	    \and
        Mart\'in Molina\thanksref{e3,addr1} 
        \and
         Marco Olivares\thanksref{e4,addr2}%etc.
}

%\thankstext[$\star$]{t1}{Thanks to the title}
\thankstext{e1}{e-mail: jose.villanueva@uv.cl}
\thankstext{e2}{e-mail: fjtapiam@gmail.com}
\thankstext{e3}{e-mail: martin.molina@alumnos.uv.cl}
\thankstext{e4}{e-mail: marco.olivaresr@mail.udp.cl}

\institute{Instituto de F\'{\i}sica y Astronom\'ia, 
	Universidad de Valpara\'iso, Avenida Gran Breta\~na 1111, Playa Ancha,
	Valpara\'iso, Chile.\label{addr1}
	\and 
	Facultad de Ingenier\'ia, Universidad Diego Portales, Avenida Ej\'ercito Libertador 441, Casilla 298--V,
	Santiago, Chile.\label{addr2}
}

%\date{Received: XX XXXX 2017 / Accepted: XX XXXX 2017}

\maketitle

\abstract{The motion of massless particles on the background of a toroidal topological black hole is analyzed in the context of conformal Weyl gravity. Null geodesics, in terms of the Jacobi elliptic functions, are found analytically. In addition, the Sagnac effect in this space-time is characterized, and we find a strong condition in the theory's parameters that is required for its existence.
}

\tableofcontents

\section{Introduction: Conformal Weyl gravity and null geodesics}
Conformal Weyl gravity (CWG) was born of an attempt to unify gravity and electromagnetism based on the principle of local invariance of a manifold, described by the metric $g_{\mu \nu}(x)$, under the change 
\begin{equation}
\label{f1} g_{\mu \nu}(x)\rightarrow\Omega^2(x)\,g_{\mu \nu}(x),
\end{equation}where $\Omega(x)$ is a smooth, strictly positive function \cite{W17,W18A,W18B,B21}.
The CWG theory can be obtained from the conformally invariant action
\begin{equation}
\label{action}
I_W=2\,\alpha_w \, \int {\rm d}^4 x\,\sqrt{-g}\,
\left[ R_{\mu \nu}\,R^{\mu \nu}-\frac{1}{3} \left(R^{\mu }_{\mu}\right)^2\right],
\end{equation}
where $\alpha_w$ is a dimensionless parameter chosen to be positive if (\ref{action}) is a positive
definite Euclidean action. The vacuum field equations associated
with this action are solved by the static, spherically symmetric
line element given by \cite{riegert84,MK89,MK91,MK91B,MK94}
\begin{equation}
{\rm d}\tilde{s}^{2}=-B(\tilde{r})\,{\rm d}\tilde{t}^{2}+\frac{{\rm d}\tilde{r}^{2}}{B(\tilde{r})}+
\tilde{r}^{2}({\rm d}\tilde{\theta}^{2}+\sin^{2}\tilde{\theta}\, {\rm d}\tilde{\phi}^{2}), \label{metrweyl}
\end{equation}
where the coordinates are defined in the range $-\infty < \tilde{t} < \infty$,
$\tilde{r}\geq0$, $0\leq\tilde{\theta}\leq\pi$, $0\leq\tilde{\phi}\leq 2\pi$, and
the lapse function $B(\tilde{r})$
is given by
\begin{equation}
B(\tilde{r})=1-\frac{\tilde{\beta}\,(2-3\tilde{\beta}\,\tilde{\gamma})}{\tilde{r}}-3\tilde{\beta}\,\tilde{\gamma}+\tilde{\gamma} \tilde{r}
- \tilde{k} \tilde{r}^{2}.
\label{lpasweyl}
\end{equation}

Here $\tilde{\beta}$, $\tilde{k}$ and $\tilde{\gamma}$ are positive constants associated with the central mass, cosmological constant and the measurements of the departure of the Weyl theory from the Einstein - de Sitter, respectively. 
Clearly, taking the limit $\tilde{\gamma}=0=\tilde{k}$ recovers the Schwarzschild case so that we can identify $\tilde{\beta} = M$.

A study of the basis and properties, together with applications of the motion of massive and massless particles in this geometry can be found, for example, in \cite{edery98,PI04,PII04,sultana10,sultana12,vo13,said13,said,Lu12,Payandeh:2012mj},
and can be obtained using the standard Lagrange procedure \cite{chandra,COV05,shutz,OSLV11,jaklitsch,VSOC13,Halilsoy:zva,LSV11}, which allows a Lagrangian $\mathcal{L}$ to be associated with the metric and then, the equation of motion to be obtained from the Lagrange's equations
\begin{equation} 
\dot{\Pi}_{q} - \frac{\partial \mathcal{L}}{\partial q} = 0,
\label{lageq} \end{equation}
where $\Pi_{q} = \partial \mathcal{L}/\partial \dot{q}$ are the conjugate momenta to the coordinate $q$, and the dot denotes a derivative with respect to the affine parameter $\tau$ along the geodesic.
Thus, in Sec. \ref{TTSEc}, and following the procedure performed  by Klemm \cite{klemm98}, we perform analytical continuations to obtain a non-trivial topology associated with toroidal topological black holes coming from CWG. In particular, we focus on the toroidal AdS black hole. Other studies associated with topological black holes can be found, for example, in Refs. \cite{olivera,ligeia}, among others. Then we obtain the conserved quantities together with the equations of motion for massless particles on these manifolds. In Sec. \ref{RMTT} the radial motion is analyzed for photons going to spatial infinite or to the singularity, while Sec. \ref{AMTT} is devoted to obtaining analytically the trajectory for photons with non-zero angular momentum, for which we employ an analysis in terms of Jacobi elliptic functions. In Sec. \ref{SETT} we apply the methods outlined by Sakurai, Tartaglia, Rizzi \& Ruggiero, among others, to obtain an analogy to the Aharanov-Bohm effect to describe the Sagnac effect for this space-time. Finally, in Sec. \ref{STT} we conclude and summarize our results.

\section{Toroidal topology}\label{TTSEc}

In order to obtain a toroidal topological black hole, we perform the following analytical continuation of the metric (\ref{metrweyl}):
\begin{eqnarray} \nonumber
&&\tilde{t} \rightarrow \sqrt{\alpha}\,t, \quad  \tilde{r}\rightarrow \frac{r}{\sqrt{\alpha}},\quad \tilde{\phi}\rightarrow \phi,\\ \nonumber
&& \tilde{\theta}\rightarrow \sqrt{\alpha}\,\theta, \quad \tilde{\beta} \rightarrow \frac{\beta}{\sqrt{\alpha}}, \quad \tilde{\gamma} \rightarrow \frac{\gamma}{\sqrt{\alpha}},\quad \tilde{k} \rightarrow k.
\end{eqnarray}
Then, by taking the limit $\alpha \rightarrow 0$, the line element becomes
\begin{equation}
\label{metrtoroidal}
{\rm d}s^2=-B(r)\,{\rm d}t^{2}+\frac{{\rm d}r^{2}}{B(r)}+r^{2}({\rm d}\theta^{2}+\theta^{2}\,{\rm d}\phi^{2}),
\end{equation}
with the lapse function
\begin{equation}
\label{laptoroidal}
B(r)=\frac{3\,\beta^2\,\gamma}{r}-3\,\beta\gamma+\gamma\,r-k\,r^2.
\end{equation}

In this case, it is possible to prove that the metric induced on the spacelike surface of constant $t$ and $r$ corresponds to a compact orientable surface with genus $g=1$, i.e., a torus, so the topology of this four-dimensional manifold becomes $\mathbb{R}^2\times S^1\times S^1$ \cite{klemm98}. 
Therefore, performing $\gamma=-2\eta/L$, $\beta=\sqrt{L/3}$, $k=-1/\ell^2$, and then evaluating at the limit $L\rightarrow\infty$, the lapse function becomes
\begin{equation}
\label{lapsads}B(r)=-\frac{2\eta}{r}+\frac{r^2}{\ell^2},
\end{equation}which for $\eta>0$ describes the AdS uncharged static toroidal black hole \cite{lemos,huang,mann97,brill,vanzo,maeda,astorino} with the event horizon placed at 
\begin{equation}
\label{horiz}r_+=(2\,\eta\, \ell^2)^{1/3}.\end{equation}
The vanished Lagrangian associated with the photons that move on this manifold can be expressed as
\begin{equation}
\label{lagr}\mathcal{L}=-\frac{1}{2} B(r) \dot{t}^2+\frac{1}{2}\frac{\dot{r}^2}{B(r)}+\frac{1}{2}r^2 (\dot{\theta}^2+\theta^2\,\dot{\phi}^2)=0.
\end{equation}

Conversely, the toroidal metric given by Eqs. (\ref{metrtoroidal}) and (\ref{lapsads}) admits the following Killing vectors field:

\begin{itemize}
	\item the {\it time-like Killing vector} $\chi=\partial_t$ is related to the {\rm stationarity} of the metric. The conserved quantity is given by
	\begin{equation}
	\label{energy}
	g_{\mu \nu}\,\chi^{\mu}\,u^{\nu}=-B(r)\,\dot{t}=-\sqrt{E}
	\end{equation}
	where $E$ is a constant of motion that cannot be
	associated with the total energy of the test particle
	because this metric is not asymptotically flat.
	\item the most general \textit{space-like Killing vector}
	is given by
	\begin{equation}
	\vec{\chi}= \left(A\,\cos \phi+B\,\sin \phi \right)\,\partial_{\theta}
	+\left[C-A\frac{\,\sin \phi}{\theta}+B\,\frac{\cos \phi}{\theta}\right]\,\partial_{\phi},
	\end{equation}
	where $A$, $B$ and $C$ are arbitrary constants.
	Is easy to see that it is a linear combination of the three
	Killing vectors
	\begin{eqnarray}\nonumber
	&&\chi_1=\partial_{\phi},\\ \nonumber
	&&\chi_2= \cos \phi\,
	\partial_{\theta}-\frac{\sin \phi}{\theta}\,\partial_{\phi},\\ \nonumber
	&&
	\chi_3=\sin \phi\,
	\partial_{\theta}+\frac{\cos \phi}{\theta} \partial_{\phi}
	\end{eqnarray} which are 
	the angular momentum operators for this space-time. The conserved quantities
	are given by
	\begin{eqnarray}\label{cca0}
	g_{\alpha \beta }\,\chi_{1}^{\alpha}\,u^{\beta}
	&=&r^{2}\,\theta^2\,\dot{\phi}
	=L_1, \\ \label{cca1}
	g_{\alpha \beta }\,\chi_{2}^{\alpha}\,u^{\beta}
	&=&r^{2}\,(\cos\phi\,\dot{\theta}-\theta\,\sin\phi\,\dot{\phi})
	=L_2, \\ \label{cca2}
	g_{\alpha \beta }\,\chi_{3}^{\alpha}\,u^{\beta}
	&=&
	r^{2}\,(\sin\phi\,\dot{\theta}+\theta\, \cos\phi\,\dot{\phi})
	=L_3,
	\end{eqnarray}
	where $L_1$, $L_2$ and $L_3$ are constants associated with the angular momentum of the particles. 
\end{itemize}

It is interesting to point out that Eqs. (\ref{cca1}) and (\ref{cca2}) implies that
\begin{equation}
\label{defj}J^2\equiv L_2^2+L_3^2=r^4 (\dot{\theta}^2+\theta^2\,\dot{\phi}^2),
\end{equation}such that,  if we focus our attention on the invariant plane $\theta=\theta_0$, so $\dot{\theta}=0$, we get that $L_1=\theta_0 \,J$. Thus, using  Eqs. (\ref{energy}) and (\ref{defj}) in Eq. (\ref{lagr}) we obtain the radial equation of motion corresponding to the one-dimensional problem
\begin{equation}
\dot{r}^{2}=E-\mathcal{V}(r),
\label{lagrhyp}
\end{equation}
where $\mathcal{V}(r)$ is the {effective potential} defined as
\begin{equation}
\label{effpot}
\mathcal{V}(r)=J^2\,\frac{B(r)}{r^2}\equiv J^2 \,V(r).
\end{equation} Here $V(r)=B(r)/r^2$ is the {\it effective potential per unit of $J^2$}.

On the other hand, without lack of generality we choose  $\theta_0=1$ and combining Eqs (\ref{defj}) and (\ref{lagrhyp}), we obtain the angular motion equation
\begin{equation}
\left(\frac{1}{r^2}\frac{dr}{d\phi}\right)^{2}= \frac{1}{b^2}-V(r)=\frac{2\,\eta}{r^3}-\left(\frac{1}{\ell^2}-\frac{1}{b^2}\right),
\label{radialpolareq}
\end{equation}
where $b=J/\sqrt{E}$ is the impact parameter.
\begin{figure}[h!]
	\includegraphics[width=.45\textwidth]{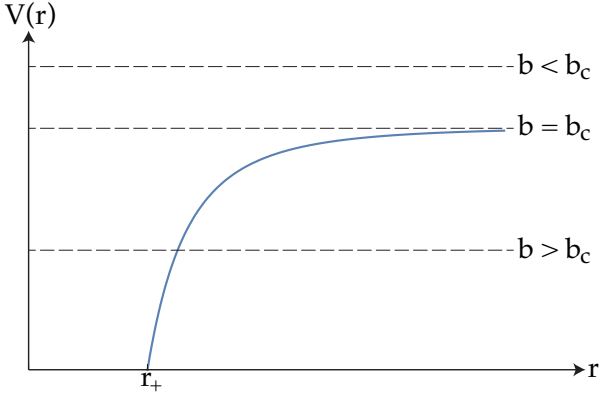}
	\caption{Graphs of the topological toroidal effective potential as a function of the radial coordinate. The critical impact parameter $b_c$ corresponds to the value of $b$ at $r\rightarrow \infty$.}\label{pot}
\end{figure}
In Fig. \ref{pot} we have plotted the effective potential (per unit of $J^2$) as a function of the radial coordinate. 
In the next sections the motion of massless particles will be analyzed analytically by integration of the equations of motion.

\section{Radial Motion}\label{RMTT}

\begin{figure}[!h]
	\begin{center}
		\includegraphics[width=80mm]{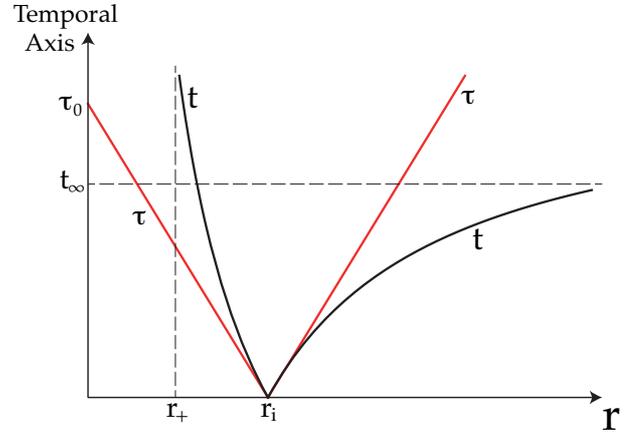}
	\end{center}
	\caption{Temporal null geodesics for radial motion described by Eqs. (\ref{tau}) and (\ref{t}). This shows some equivalence with its spherical counterpart previously studied by Cruz et al. \cite{COV05} for photons moving to $r_+$. The novel result appears for photons moving to infinite in which, as is measured by an observer in $r_i$, it takes a finite coordinate time $t_{\infty}$ to reach infinite. This feature was reported before by Villanueva \& V\'asquez but in the context of Lifshitz space-times \cite{Villanueva:2013gra}.}
	\label{temp}
\end{figure}

Radial motion corresponds to a trajectory with null angular momentum $J=0$, so photons fall toward the singularity or to the spatial infinite. From Eq. (\ref{effpot}) we can see that photons in radial motion have a null effective potential $\mathcal{V}(r)=0$, so that Eqs. (\ref{energy}) and (\ref{lagrhyp}) become
\begin{equation}
\frac{dr}{d\tau}=\pm \sqrt{E},
\label{mr.1}
\end{equation}
and
\begin{equation}
\frac{\textrm{d}r}{dt}=\pm B(r)=\pm\frac{1}{\ell^2}\left(\frac{r^3-r_+^3}{r}\right),
\label{mr.2}
\end{equation}where the sign $-$ ($+$) corresponds to photons  falling to the event horizon (spatial infinite). Assuming that $t=\tau=0$ at $r=r_i$, then a straightforward integration of  Eq. (\ref{mr.1}) leads to
\begin{equation}
\label{tau}\tau(r)=\pm\frac{r-r_i}{\sqrt{E}},
\end{equation}while an integration of Eq. (\ref{mr.2}) becomes
\begin{eqnarray}
\nonumber t(r)=&&\pm \frac{\ell^2}{\sqrt{3}\,r_+}\left\{\arctan\left(\frac{2r+r_+}{\sqrt{3}\,r_+}\right)-\arctan\left(\frac{2r_i+r_+}{\sqrt{3}\,r_+}\right)+\right.\\ \label{t}
&&\left.+\log\left[\frac{r_i^3-r_+^3}{r^3-r_+^3}\left(\frac{r-r_+}{r_i-r_+}\right)^3 \right]\right\}.
\end{eqnarray}
Obviously, in the proper system photons cross the event horizon in a finite time $\tau(r_+)\equiv\tau_+=(r_i-r_+)/\sqrt{E}$ and, eventually, arrive to the singularity in a finite time $\tau(0)\equiv\tau_0=r_i/\sqrt{E}$. Also, they eternally approach the spatial infinite i. e., $\tau(\infty)\rightarrow \infty$. On the other hand, an observer at $r_i$ sees that photons take an infinite coordinate-time even to arrive at $r_+$, while it takes a finite coordinate-time even to arrive at the spatial infinite, given by
\begin{equation}
\label{tinf}t_{\infty}=\frac{\ell^2}{\sqrt{3}\,r_+}\left\{\frac{\pi}{2}-\arctan\left(\frac{2r_i+r_+}{\sqrt{3}\,r_+}\right)+\log\left[\frac{r_i^3-r_+^3}{(r_i-r_+)^3} \right]\right\}.
\end{equation}The existence of this time is due to the presence of the cosmological term on the toroidal topology and depends on the position of the observer $r_i$. A similar feature was reported before by Villanueva \& V\'asquez, but in the context of Lifshitz black holes \cite{Villanueva:2013gra}. The behavior of both proper and coordinate time is shown in Fig. \ref{temp}.

\section{Angular Motion}\label{AMTT}

\begin{figure}[!h]
	\begin{center}
		\includegraphics[width=83mm]{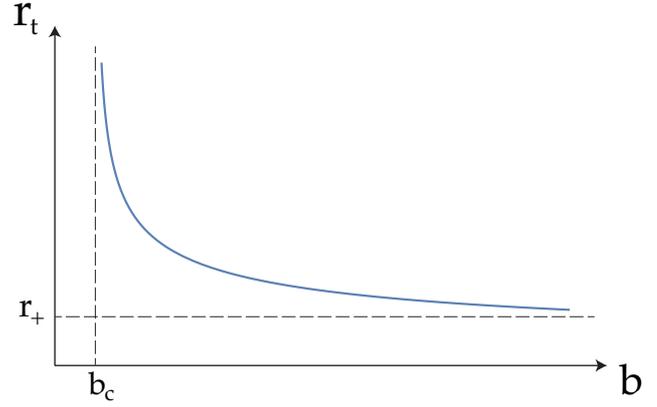}
	\end{center}
	\caption{Plot for the turning point or {\it apoastron}, $r_t$ as a function of the impact parameter $b$. The validity of this function is in $b_c<b<\infty$.}
	\label{rtfig}
\end{figure}

\begin{figure}[!h]
	\begin{center}
		\includegraphics[width=83mm]{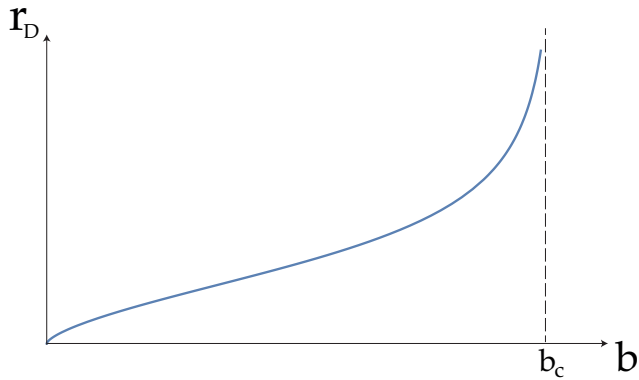}
	\end{center}
	\caption{Plot for the modulus of the real negative root $r_D$ as a function of the impact parameter associated with the unbounded motion of photons, i.e. $0<b<b_c$.}
	\label{rdfig}
\end{figure}

This section is devoted to studying the angular motion of the test particles ($J\neq 0$), which depends on the value of the impact parameter $b$. From Fig. \ref{pot} we can see that there are two distinct regions separated by the critical impact parameter $b_c=\ell$. Thus, if $b>b_c$ the motion will be confined with a turning point, {\it the apoastron distance} $r_t$, placed at
\begin{equation}
\label{rt}r_t(b)=\frac{r_+}{\left[1-\left(\frac{b_c}{b}\right)^2\right]^{1/3}},
\end{equation}whereas if $b<b_c$ does not exist the turning point, and the motion is unbound with a negative characteristic distance (without physical meaning), the magnitude of which is
\begin{equation}r_D(b)=\frac{r_+}{\left[\left(\frac{b_c}{b}\right)^2-1\right]^{1/3}}. \label{rd}\end{equation}
Both distances, $r_t$ and $r_D$, play an important role in the determination of trajectories because they depend strongly on the impact parameter $b$ (see Figs. \ref{rtfig} and \ref{rdfig}).

\subsection{Confined motion}
\begin{figure}[!h]
	\begin{center}
		\includegraphics[width=75mm]{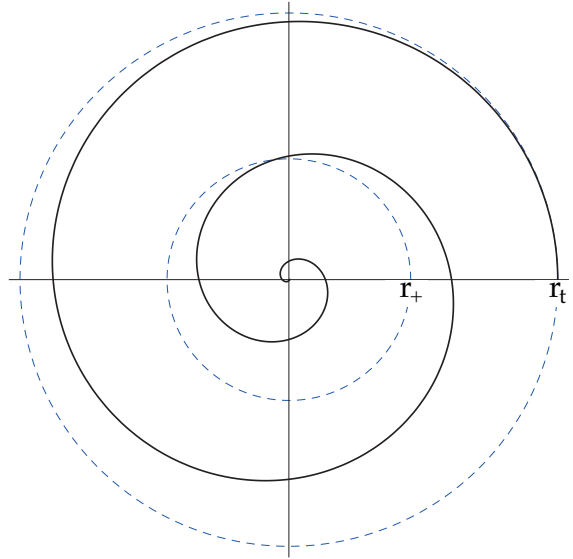}
	\end{center}
	\caption{Polar plot for a confined trajectory of photons from a distance $r_t$.}
	\label{conf}
\end{figure}

Returning to the general Eq. (\ref{radialpolareq}), we first consider the case 
when the impact parameter lies between $b_c<b<\infty$, so using the variable $u=1/r$ with $u_t=1/r_t$, we can write
\begin{eqnarray}
\nonumber \frac{{\rm d}u}{{\rm d}\phi}&=&\pm \sqrt{2\,\eta}\,\sqrt{u^3-u_t^3}\\\label{raduq}
&=&\pm \sqrt{2\,\eta}\,\sqrt{(u-u_t)\left[\left(u+\frac{u_t}{2}\right)^2+\frac{3\, u_t^2}{4}\right]}.
\end{eqnarray}Using the substitution
\begin{equation}
\label{subs1}u=u_t\,\left(\sqrt{3}\,\cot^2 \frac{\xi}{2}+1\right),
\end{equation} where the range of $u$ is $u_t\leq u<\infty$, and the corresponding range of $\xi$ is $\xi_t=\pi\leq \xi<2\pi$, Eq. (\ref{raduq}) can be reduced to the elliptic form
\begin{eqnarray}
\nonumber& \pm&  \alpha_1\,\int_{\phi_t}^{\phi}{\rm d}\phi'=\int_{\pi}^{\xi}\frac{{\rm d}\varphi}{\sqrt{1-k_1^2\,\sin^2 \varphi}}\\\label{phiu}
&=&\int_{0}^{\xi}\frac{{\rm d}\varphi}{\sqrt{1-k_1^2\,\sin^2 \varphi}}-\int^{\pi}_{0}\frac{{\rm d}\varphi}{\sqrt{1-k_1^2\,\sin^2 \varphi}},
\end{eqnarray}where
\begin{equation}
\label{modk}\alpha_1= \sqrt{2\sqrt{3}\,\eta\,u_t},\qquad k_1=\sqrt{\frac{2-\sqrt{3}}{4}}.
\end{equation}
So, we may write the solution for $\phi$ as
\begin{equation}
\label{solthet}\pm \alpha_1\, \phi=2K(k_1)-F(\xi, k_1),
\end{equation}where $K(k)$ and $F(\psi, k)$ are the complete and incomplete elliptic integrals of the first kind, respectively, and we have assumed that $\phi_t=0$. Therefore, by using some formulas and identities of the Jacobian elliptic functions, we can write the equation of the trajectory as
\begin{equation}
\label{trayrt}r(\phi, b)=\frac{r_t(b)}{1+\sqrt{3}\,\textrm{tn}^2\left(\frac{\alpha_1(b)\,\phi}{2}, k_1\right)\,\textrm{dn}^2\left(\frac{\alpha_1(b)\,\phi}{2}, k_1\right)},
\end{equation}where tn$(x, k)$ and  dn$(x, k)$  are Jacobi's elliptic functions (see \ref{app:jef} and Refs. \cite{byrd,hancock,Armitage}). Obviously, this trajectory depends on the impact parameter and is shown in Fig. \ref{conf} for photons falling from $r_t$.

\subsection{Unbounded motion}
\begin{figure}[!h]
	\begin{center}
		\includegraphics[width=75mm]{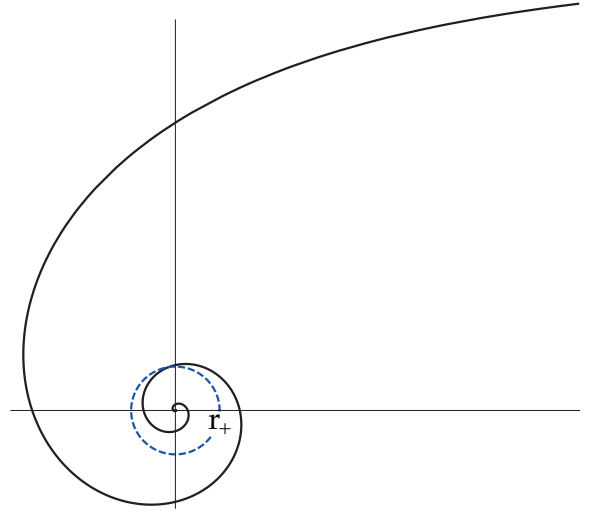}
	\end{center}
	\caption{Polar plot for the unbounded trajectory, Eq. (\ref{radunb}), followed by photons whose impact parameter lies between $0<b<b_c$.}
	\label{unbfig}
\end{figure}

We now consider the case when the impact parameter lies between $0<b<b_c$, which means that the real root is a negative one $r_n\equiv-r_D$, where $r_D$ is given by Eq. (\ref{rd}), so there is no turning point. Obviously, this negative root lacks physical interpretation, but it is important for determining the shape of the trajectory. Therefore, making $u=1/r$ again with $u_D=1/r_D$, the Eq. of motion (\ref{radialpolareq}) can be written as
\begin{eqnarray}
\nonumber \frac{{\rm d}u}{{\rm d}\phi}&=&\pm \sqrt{2\,\eta}\,\sqrt{u^3+u_D^3}\\\label{unbound}
&=&\pm \sqrt{2\,\eta}\,\sqrt{(u+u_D)\left[\left(u-\frac{u_D}{2}\right)^2+\frac{3\, u_D^2}{4}\right]}.
\end{eqnarray}
In order to integrate Eq. (\ref{unbound}), we now make the substitution
\begin{equation}
\label{cv2}u=u_D\left(\sqrt{3}\,\cot^2 \frac{\xi}{2}-1\right),
\end{equation}where the range is
\begin{eqnarray}
\label{rang1} &&u\rightarrow 0\quad ({\rm infinity}), \quad \xi_{\infty}=\arccos\left(-\frac{2-\sqrt{3}}{2}\right),\\
\label{rang2}&&u\rightarrow \infty \quad ({\rm singularity}), \quad \xi_{s}=\pi,
\end{eqnarray}
so the quadrature becomes
\begin{eqnarray}
\nonumber &&\pm  \alpha_2\,\int_{\phi_{\infty}}^{\phi}{\rm d}\phi'=\int_{\xi_{\infty}}^{\xi}\frac{{\rm d}\varphi}{\sqrt{1-k_2^2\,\sin^2 \varphi}}\\\label{phiun}
&=&\int_{0}^{\xi}\frac{{\rm d}\varphi}{\sqrt{1-k_2^2\,\sin^2 \varphi}}-\int^{\xi_{\infty}}_{0}\frac{{\rm d}\varphi}{\sqrt{1-k_2^2\,\sin^2 \varphi}},
\end{eqnarray}with
\begin{equation}
\label{modku}\alpha_2= \sqrt{2\sqrt{3}\,\eta\,u_D},\qquad k_2=\sqrt{\frac{2+\sqrt{3}}{4}}.
\end{equation}Note from Eqs. (\ref{modk}) and (\ref{modku}) that the module of one trajectory corresponds to the complementary module of the other, $k_1=\sqrt{1-k_2^2}=k'_2$ and $k_2=\sqrt{1-k_1^2}=k'_1$. Therefore, assuming that $\phi_{\infty}=0$ we may write
\begin{equation}
\label{unbthe}\pm \alpha_2\,\phi=F(\xi, k_2)-F(\xi_{\infty}, k_2),
\end{equation}which implies that the trajectory now is described by the polar equation
\begin{equation}
\label{radunb}r(\phi, b)=\frac{r_D(b)}{\sqrt{3}\,{\rm cs}^2(\Theta, k_2) \,{\rm nd}^2(\Theta, k_2)-1},\end{equation}
where cs$(x, k)$ and nd$(x, k)$ are Jacobi's elliptic functions (see \ref{app:jef}), and the phase $\Theta$ is given by
\begin{equation}
\label{phas}\Theta=\frac{F(\xi_{\infty}, k_2)-\alpha_2(b)\,\phi}{2}.
\end{equation}In Fig. \ref{unbfig} we have plotted the unbounded trajectory (\ref{radunb}) for photons coming from infinite.

\section{The Sagnac effect}\label{SETT}
In this section we describe  the Sagnac effect \cite{sagnac} by applying the formalism developed by Sakurai \cite{sakurai}, Tartaglia \cite{tartaglia}, Rizzi \& Ruggiero \cite{rizzirugg04,Rizzi:2003bf,Rizzi:2003uc,ruggiero05}, among others, to the exterior space-time of the topoloical toroidal black hole. This approach was used to calculate this effect in the exterior space-time to an uncharged spherical symmetric black hole in Conformal Weyl gravity \cite{sultana14}. For the non-rotating toroidal metric given by Eqs. (\ref{metrtoroidal}) and (\ref{lapsads}) written in the usual Schwarzschild--like coordinates ($ct', r', \theta', \phi'$):
\begin{eqnarray}
\nonumber
{\rm d}s^2&=&-\left(-\frac{2\eta}{r'}+\frac{r'^2}{\ell^2}\right)\,c^2{\rm d}t'^{2}+\frac{{\rm d}r'^{2}}{\left(-\frac{2\eta}{r'}+\frac{r'^2}{\ell^2}\right)}+\\\label{metrtoroidal2}
&&+r'^{2}({\rm d}\theta'^{2}+\theta'^{2}\,{\rm d}\phi'^{2}),
\end{eqnarray}so the transformation to the azimuthal frame of the rotating platform
\begin{equation}
\label{plat}ct=ct',\quad r=r',\quad \theta=\theta',\quad \phi=\phi'-\Omega t',
\end{equation}where $\Omega$ is the constant angular velocity of the physical frame, yields the metric (after setting $r=R$ and $\theta=1$)
\begin{eqnarray}
\nonumber
{\rm d}s^2&=&-\left(-\frac{2\eta}{R}+\frac{R^2}{\ell^2}-\frac{\Omega^2 R^2}{c^2}\right)\,{\rm d}(ct)^{2}+R^{2}{\rm d}\phi^{2}+\\\label{metrtoroidal3}
&&+2\frac{\Omega}{c}\,R^2\,{\rm d}\phi\,{\rm d}(ct).
\end{eqnarray}
Therefore, the non-zero components of the unit vector $\gamma^{\alpha}$ along the trajectory $r=R$ are given by 
\begin{eqnarray}
\label{gammafunc1} \gamma^t&=&\frac{1}{\sqrt{-g_{tt}}}=\gamma_J,\\
\label{gammafunc2} \gamma_t&=&-\sqrt{-g_{tt}}=-\gamma_J^{-1},\\
\label{gammafunc3} \gamma_{\phi}&=&g_{\phi t}\,\gamma^t=\frac{\Omega}{c}\,R^2\,\gamma_J,
\end{eqnarray}
where
\begin{equation}
\label{gaj}\gamma_J=\left[R^2\left(\frac{1}{\ell^2}-\frac{\Omega^2}{c^2}\right)-\frac{2\eta}{R}\right]^{-1/2}.
\end{equation}In terms of this component, the gravito-electric and gravito-magnetic potentials are given by
\begin{eqnarray}
\label{pot1}\phi^G&=&-c^2\,\gamma^t=-c^2\,\gamma_J,\\
\label{pot2}A_{\phi}^G&=&c^2 \frac{\gamma_{\phi}}{\gamma_t}=- c\,\Omega\, R^2\, \gamma_J^2.
\end{eqnarray}As was shown in \cite{Rizzi:2003uc}, it is possible to express the phase shift and time delay between light beams detected by a co-moving observer on the interferometer in terms of the gravito-magnetic induction field,  $\vec{B}^G=\vec{\nabla}\times \vec{A}^G$, by means of the expressions
\begin{equation}
\label{delph}\Delta \Phi=\frac{2\, \epsilon\, \gamma_t}{\hbar\,c^3}\int_{S}\vec{B}^G \cdot {\rm d}\vec{S}=\frac{2\, \epsilon\, \gamma_t}{\hbar\,c^3}\int_{\zeta(S)}\vec{A}^G \cdot {\rm d}\vec{r},
\end{equation}and
\begin{equation}
\label{delt}\Delta \tau=\frac{2\gamma_t}{c^3}\int_{S}\vec{B}^G \cdot {\rm d}\vec{S}=\frac{2 \gamma_t}{c^3}\int_{\zeta(S)}\vec{A}^G \cdot {\rm d}\vec{r},
\end{equation}where $\epsilon$ is the relative energy of the photon as measured in the interferometer.
\begin{figure}[!h]
	\begin{center}
		\includegraphics[width=85mm]{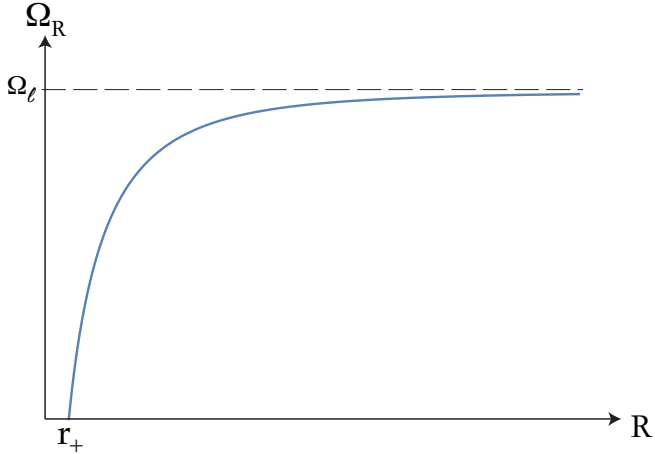}
	\end{center}
	\caption{Plot for the maximum angular velocity $\Omega_R$ as a function of the radius of the orbit $R$ for an interferometer in the Sagnac effect.}
	\label{om}
\end{figure}
Therefore, the phase shift turns out to be
\begin{equation}
\label{shift1}\Delta \Phi=\frac{4\pi \epsilon}{\hbar c}\frac{\tilde{\Omega}R}{\sqrt{1-\tilde{\Omega}^2-\frac{r_+^3}{R^3}}},
\end{equation}while the time delay turns out to be
\begin{equation}
\label{shift2}\Delta \tau=\frac{4\pi }{c}\frac{\tilde{\Omega}R}{\sqrt{1-\tilde{\Omega}^2-\frac{r_+^3}{R^3}}},
\end{equation}where $\tilde{\Omega}=\Omega/\Omega_{\ell}$ is the dimensionless angular velocity, and $\Omega_{\ell}\equiv c/\ell$. Thus, from this last equation, we can see that if $\Omega=0$, i.e., performing a measure of the propagation time in a non-rotating frame, no Sagnac effect arises. Also, there is an upper limit for the angular velocity $\Omega_R$ related to the  radius of the orbit $R$ (see Fig. \ref{om}) given by
\begin{equation}
\label{condrw}\Omega<\Omega_R \equiv \Omega_{\ell}\,\sqrt{1-\left(\frac{r_+}{R}\right)^3}.
\end{equation}

\section{Summary}\label{STT}

In this paper we have studied the null structure of the geodesics for a toroidal topological space-time which surrounds a black hole in the conformal Weyl gravity. First, we obtained an explicit behavior of radial photons to conclude that, while no changes in the motion to the singularity with respect to the Schwarzschild anti-de Sitter counterpart is found, there is a non-trivial coordinate time $t_{\infty}$ for the description of the motion to the spatial infinite (see Eq. (\ref{tinf})). A similar result was obtained by Villanueva \& V\'asquez but in the context of the Lifshitz space-time \cite{Villanueva:2013gra}. Next, following the standard Lagrangian procedure, we have obtained analytically the trajectories of the confined and unbounded angular motion for photons in terms of Jacobi elliptic functions, Eqs. (\ref{trayrt}) and (\ref{radunb}), and then we have shown our results in Fig. \ref{conf} and Fig. \ref{unbfig}, respectively. Obviously, these trajectories depend on the impact parameter $b$ and, due to the topology, always fall to the singularity, which is a characteristic of AdS space-times.
Finally, the Sagnac  effect has been studied for this topological space-time. Our result is consistent with those obtained previously in other geometries in the sense that no Sagnac effect arises for a non-rotating frame. In addition, we have found a strong condition  for its existence, which depends on the theory's parameters $\{\eta, \ell\}$ (in $r_+$ and $\Omega_{\ell}$) as well as on the radius of the circular orbit $R$. This condition is given in Eq. (\ref{condrw}), c.f.
\begin{equation}
\nonumber \Omega<\Omega_R \equiv \Omega_{\ell}\,\sqrt{1-\left(\frac{r_+}{R}\right)^3},
\end{equation}for which the upper limit for the angular velocity $\Omega_R$ was plotted in Fig. \ref{om} as a function of the radius $R$.

Finally, our study provides a simple physical visualization of the null trajectories and their main issues, and complements other studies carried out in the standard and/or trivial topology \cite{vo13,sultana14}.

\appendix
\section{ A brief review of Jacobian elliptic functions}\label{app:jef}
As a starting point, let us consider
the elliptic integral \cite{byrd,hancock,Armitage}
\begin{eqnarray}\nonumber
u(y, k)\equiv u&=&\int_0^y\frac{dt}{\sqrt{(1-t^2)(1-k^2\,t^2)}}\\\label{a1}
&=&
\int_0^{\varphi}\frac{d\theta}{\sqrt{1-k^2\sin^2\theta}}=
F(\varphi, k),
\end{eqnarray}
where $F(\varphi, k)$ is the {\it normal elliptic integral of the first kind},
and $k$ is the {\it modulus}.
The problem of the inversion of this integral was
studied and solved by Abel and Jacobi, and leads
to the inverse function defined by
$y=\sin\varphi=\textrm{sn}(u, k)$
with $\varphi=\textrm{am}\, u$, and are called
{\it Jacobi elliptic sine} $u$ and {\it amplitude} $u$.

The function sn $u$ is an odd elliptic function of order two. It possesses
a simple pole of residue $1/k$ at every point congruent to
$i K'$ (mod $4K$, 2$i K'$) and a simple pole of residue
$-1/k$ at points congruent to $2K+i K'$ (mod 4 $K$, $2 i K'$),
where $K\equiv K(k)=F(\pi/2, k)$ is the {\it complete elliptic integral of the first kind}, $K'=F(\pi/2, k')$, and $k'=\sqrt{1-k^2}$ is the {\it complementary modulus}.

Two other functions can then be defined by cn$(u, k) = \sqrt{1-y^2} = \cos \varphi$, which is called the {\it Jacobi elliptic cosine} $u$, and is an even function of order two; $\textrm{dn}(u, k)=\sqrt{1-k^2\,y^2}=\Delta \varphi=\sqrt{1-k^2\,\sin \varphi}$, called the {\it Jacobi elliptic delta} $u$, which is an even function.
The set of functions $\{\textrm{sn}\, u,\, \textrm{cn}\, u,\, \textrm{dn}\, u\}$
are called {\it Jacobian elliptic functions}, and 
take the following special values:
\begin{eqnarray}
&&\textrm{sn} (u, 0)=\sin u,\quad \textrm{sn} (u, 1)=\tanh u,\\
&&\textrm{cn} (u, 0)=\cos u,  \quad \textrm{cn} (u, 1)=\textrm{sech}\, u,\\
&& \textrm{dn} (u, 0)=1,\, \qquad \,\,\textrm{dn} (u, 1)=\textrm{sech}\, u,\\
&& \textrm{tn} (u, 0)=\tan u, \quad \textrm{tn} (u, 1)=\sinh\, u.
\end{eqnarray}
The quotients and reciprocal of $\{\textrm{sn}\, u,\, \textrm{cn}\, u,\, \textrm{dn}\, u\}$
are designated in {\it Glaisher's notation} by
\begin{eqnarray}
&&\textrm{ns}\, u=\frac{1}{\textrm{sn} \, u},\quad \textrm{cs}\, u=\frac{\textrm{cn} \, u}{\textrm{sn} \, u},\quad \textrm{ds}\, u=\frac{\textrm{dn} \, u}{\textrm{sn} \, u},\\
&&\textrm{nc}\, u=\frac{1}{\textrm{cn} \, u},\quad \textrm{tn}\, u\equiv \textrm{sc}\, u=\frac{\textrm{sn} \, u}{\textrm{cn} \, u},\quad \textrm{dc}\, u=\frac{\textrm{dn} \, u}{\textrm{cn} \, u},\\
&& \textrm{nd}\, u=\frac{1}{\textrm{dn} \, u},\quad \textrm{sd}\, u=\frac{\textrm{sn} \, u}{\textrm{dn} \, u},\quad \textrm{cd}\, u=\frac{\textrm{cn} \, u}{\textrm{dn} \, u}.
\end{eqnarray}
Therefore, in all, we have twelve Jacobian elliptic functions.
Finally, some useful fundamental relations between Jacobian elliptic functions used across this work are 
\begin{eqnarray}
&&\textrm{sn}^2 u +  \textrm{cn}^2 u=1,\\
&&\textrm{dn}^2 u+ k^2\, \textrm{sn}^2 u  =1,\\
&& \textrm{dn}^2 u-k^2\,\textrm{cn}^2 u=k'^2,\\
&& \textrm{cn}^2 u+k'^2\,\textrm{sn}^2 u =\textrm{dn}^2 u,\\
&& \frac{1-\textrm{cn}\, 2u}{1+\textrm{cn}\, 2u}=\textrm{tn}^2 u\,\textrm{dn}^2 u,\\
&& \textrm{cn}\, (u\pm 2K)=-\textrm{cn}\, u.
\end{eqnarray}

%%%%%%%%%%%%%%%%%%%%%%%%%%%%%5
\begin{acknowledgement}
The authors acknowledge useful conversations with Prof. Dr. Ricardo Troncoso, Prof. Dr. Graeme Candlish and Dr. Helen Lowry. In addition, we acknowledge to {\it Centro de Astrof\'isica de Valpara\'iso} (CAV) for support part of this work.
\end{acknowledgement}

\end{document}